\documentclass[a4paper,fleqn,usenatbib,useAMS]{mnras}

\usepackage{graphicx}	
\usepackage{amsmath}	
\usepackage{amssymb}	
\usepackage{multicol}        
\usepackage{bm}		
\usepackage[T1]{fontenc}
\usepackage{ae,aecompl}
\usepackage{newtxtext,newtxmath}
\usepackage[table,dvipsnames]{xcolor}			
\usepackage[normalem]{ulem}

\defcitealias{duguid_shear_driven_2023}{Paper I}

\title[Shear-driven magnetic buoyancy]{Shear-driven magnetic buoyancy in the solar tachocline:
Dependence of the mean electromotive force on diffusivity and latitude
}
\author[C. D. Duguid et al.]{
	Craig D. Duguid$^{1,2}$\thanks{E-mail: craig.d.duguid@durham.ac.uk},
	Paul J. Bushby$^{2}$,
	and Toby S. Wood$^{2}$ 
	\\
	$^{1}$Department of Mathematical Sciences, Durham University, Upper Mountjoy Campus, Durham, DH1 3LE, UK\\
 $^{2}$School of Mathematics, Statistics and Physics, Newcastle University, Newcastle Upon Tyne, NE1 7RU, UK
}

\date{Accepted XXX. Received YYY; in original form ZZZ}
\pubyear{2022}

\begin{document}
\label{firstpage}
\pagerange{\pageref{firstpage}--\pageref{lastpage}}
\maketitle

\begin{abstract} 
The details of the dynamo process that is responsible for driving the solar magnetic activity cycle are still not fully understood. 
In particular, whilst differential rotation provides a plausible mechanism for the regeneration of the toroidal (azimuthal) component of the large-scale magnetic field, there is ongoing debate regarding the process that is responsible for regenerating the Sun's large-scale poloidal field.
Our aim is to demonstrate that magnetic buoyancy, in the presence of rotation, is capable of producing the necessary regenerative effect.
Building upon our previous work, we carry out numerical simulations of a local Cartesian model of the tachocline, consisting of a rotating, fully compressible, electrically conducting fluid with a forced shear flow.
An initially weak, vertical magnetic field is sheared into a strong, horizontal magnetic layer that becomes subject to magnetic buoyancy instability.
By increasing the Prandtl number we lessen the back reaction of the Lorentz force onto the shear flow, maintaining stronger shear and a more intense magnetic layer.
This in turn leads to a more vigorous instability and a much stronger mean electromotive force, which has the potential to significantly influence the evolution of the mean magnetic field. 
These results are only weakly dependent upon the inclination of the rotation vector, i.e.~the latitude of the local Cartesian model. 
Although further work is needed to confirm this, these results suggest that magnetic buoyancy in the tachocline is  a viable poloidal field regeneration mechanism for the solar dynamo. 
\end{abstract}

\begin{keywords} 
MHD -- Sun: magnetic fields -- dynamo -- Sun: rotation -- hydrodynamics -- instabilities
\end{keywords}


\section{Introduction}\label{section_introduction}

One of the leading paradigms for the maintenance of the Sun's observed large-scale magnetic field is the $\alpha\omega$-dynamo, following ideas introduced by \citet{parker_hydromagnetic_1955}.
This dynamo mechanism requires differential rotation to stretch magnetic field lines in the toroidal (azimuthal) direction, via the $\omega$-effect.
The dynamo loop is then closed by the $\alpha$-effect, whereby the action of rising, twisting motions (acting upon this shear-generated toroidal field) lead to the production of a mean poloidal field.
The strongest rotational shear is known to be located in the solar tachocline, at the base of the convective envelope, making it a likely site for the $\omega$-effect to be operating.
However, the exact source of the $\alpha$-effect remains controversial. 
Following Parker's original ideas \citep{parker_hydromagnetic_1955,parker_solar_1993}, the $\alpha$-effect is often attributed to cyclonic convection. 
Yet simulations of dynamos driven by (moderately) turbulent rotating convection typically produce only disordered, small-scale magnetic fields \citep[e.g.][]{cattaneo_dynamo_2006,favier_bushby_2013}, suggesting a negligible $\alpha$-effect in such systems. 
It is therefore natural to seek alternative processes that can produce a similar ``rise and twist'' regenerative effect.

It is well known that regions of strong magnetic field tend to be less dense than their surroundings \citep{parker_formation_1955,jensen_tubes_1955}, and are therefore buoyant.
Motivated by the solar tachocline, direct numerical simulations have been used to study the magnetic buoyancy instability in a horizontal magnetic layer that is induced by a vertical shear \citep{vasil_magnetic_2008,vasil_constraints_2009,silvers_interactions_2009,silvers_double-diffusive_2009,barker_magnetic_2012}.
In these studies, in which the effects of rotation are neglected, an initially weak, uniform, vertical magnetic field is wound up by a shear flow that is maintained by a fixed volumetric forcing.
Despite the similarity of the models, the results are surprisingly varied.
Whereas \citet{vasil_magnetic_2008} found magnetic buoyancy instability only with an unrealistically strong shear forcing (an order of magnitude larger than the sound speed), \citet{silvers_double-diffusive_2009} were able to excite magnetic buoyancy instability with a significantly weaker shear.
\citet{silvers_double-diffusive_2009} attributed their success to double-diffusive effects, in particular the small ratio of magnetic to thermal diffusivity.
More recently, however, \citet{lewis_fluid_2022} has pointed out that another crucial difference between these two models is in the choice of initial condition.
If the fluid is initialised from rest (as in \citealt{vasil_magnetic_2008}) then the Lorentz force from the growing magnetic field can act to suppress the shear before the field itself becomes buoyantly unstable.
By contrast, if the shear flow is established before the vertical magnetic field is introduced (as in \citealt{silvers_double-diffusive_2009}) then the toroidal field can become stronger by several orders of magnitude, greatly increasing the likelihood of magnetic buoyancy instability.
It should be emphasised that, in both of these scenarios, the instability arises as a perturbation in an evolving system, and the dominant mode of instability changes over the course of this evolution \citep{lewis_fluid_2022}.

The effects of rotation on magnetic buoyancy instability are non-trivial \citep[e.g. see the review by][]{hughes_tachocline_2007}.
However, idealised models of the magnetic buoyancy instability with an imposed magnetic layer have shown that it can produce a mean electromotive force (EMF) that could, in principle, lead to the production of a mean poloidal field \citep[e.g.][]{moffatt_magnetic_1978,davies_mean_2011}.
In our recent publication \citep[][henceforth \citetalias{duguid_shear_driven_2023}]{duguid_shear_driven_2023}, we considered the evolution of a shear-generated magnetic layer, extending the model of \citet{silvers_double-diffusive_2009} to include rotation.
For sufficiently rapid rotation, a systematic positive-signed EMF was produced in the direction parallel to the mean field.
Such a systematic mean EMF is likely to be conducive to dynamo action of $\alpha\omega$-type as per the original ideas of \citet{parker_hydromagnetic_1955}.
However, further work is needed to confirm this idea, and there are some key limitations that will need to be overcome before a successful dynamo can be produced.
In particular, the parameter regime in which these simulations operate is an extremely challenging one \citep[similar to that of][]{silvers_double-diffusive_2009}: the viscosity and magnetic diffusivity are both several orders of magnitude smaller than the thermal diffusivity, which greatly increases the range of scales that must be resolved in the simulations. 
It would not be feasible to run a global spherical dynamo calculation in this parameter regime, and even local Cartesian dynamo models will be very demanding in terms of computational requirements. 
Another limitation is that the shear flow in all of the simulations presented in \citetalias{duguid_shear_driven_2023} eventually became ``un-tachocline-like'', either through the suppression of differential rotation or through the generation of strong meridional shear, as a result of the Lorentz force associated with the large-scale magnetic field.
The simulations were therefore inherently transient, making it difficult to extract meaningful statistics.
Finally, these calculations were restricted solely to the polar regions of the solar interior, which is not the primary region of interest for the solar dynamo. 
Overcoming these limitations forms the motivation for the present paper.

The main aim of this paper is to demonstrate that many of these issues can be overcome by making appropriate changes to the parametric regime of the simulations.
In Section 2, we present brief details of the model, focusing upon the differences between this model and that considered in \citetalias{duguid_shear_driven_2023}.
This is followed, in Section 3, by a detailed numerical study, focusing upon the effects of varying the diffusive parameters. 
Section 4 describes the effect of varying the latitude of the domain, i.e.~the angle between the rotation axis and gravity.
We conclude the paper with a summary of the main findings and further discussion.


\section{Numerical model}\label{section_model}
\subsection{Equations and parameters}\label{section_model_setup}
\begin{figure}
	\begin{center}
    \includegraphics[width=0.8\columnwidth]{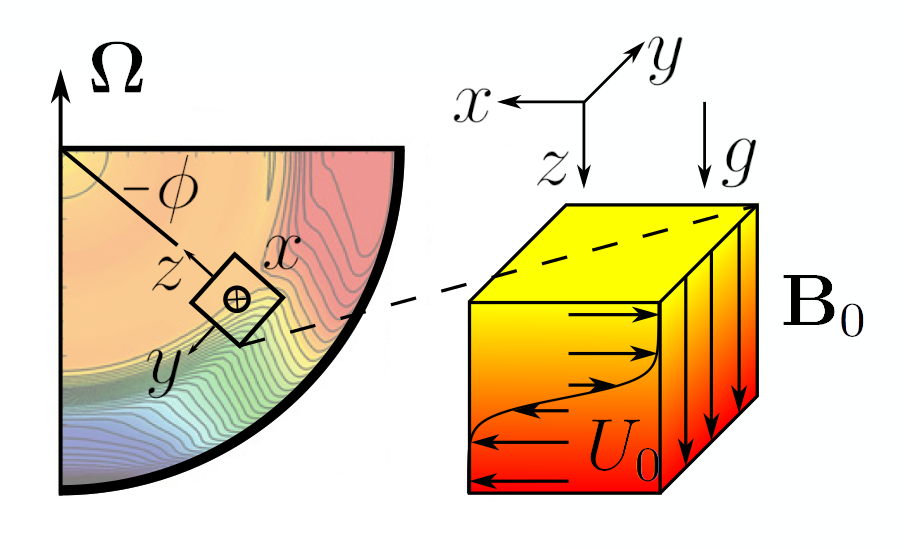}
	\caption{Schematic of the Cartesian model, which represents a local patch of the solar interior located at some latitude $\phi$ (note that negative values of $\phi$ correspond to the southern hemisphere). 
    The arrows in the $x$-direction represent the tachocline-like shear flow and the vertical field lines represent the imposed initial magnetic field $B_0\mathbf{e}_z$.}
	\label{fig_geometry}
	\end{center}
\end{figure}

The model, which we shall briefly summarise in this section, is a generalisation of that presented in \citetalias{duguid_shear_driven_2023}.
We consider a local Cartesian box within the tachocline, which we model as an ideal gas that is electrically conducting and fully compressible.
As indicated in Fig.~\ref{fig_geometry}, the local Cartesian axes $x,y,z$ are defined in the directions of increasing longitude, colatitude, and depth, respectively.
When rotation is included, we work in the frame rotating with constant angular velocity $\boldsymbol{\Omega}=-\Omega(\cos(\phi) \boldsymbol{e}_y + \sin(\phi) \boldsymbol{e}_z)$, where $\phi$ represents the latitude of the model and the scalar quantity $\Omega$ defines the rotation rate. 
Note that negative values of $\phi$ correspond to the southern hemisphere. 
A volumetric forcing is applied to the momentum equation to produce an azimuthal shear flow (in the $x$ direction), mimicking the local differential rotation within the tachocline.
Because the rotation axis is tilted, the forcing required is slightly different to that in \citetalias{duguid_shear_driven_2023}, and is defined below by equation~(\ref{maths_shear_forcing}).
The initial state is that of the hydrodynamically stable shear flow with a weak vertical magnetic field, $\boldsymbol{B} = B_0 \boldsymbol{e}_z$, as well as small-amplitude thermal noise to initiate three-dimensional motions of the fluid.
The vertical field is subsequently stretched out in the $x$-direction by the shear flow, producing a magnetic layer that eventually becomes dynamically significant and, potentially, susceptible to magnetic buoyancy instability.
The domain size is the same as in \citetalias{duguid_shear_driven_2023}, with $0\leq x \leq 2d$, $0\leq y \leq 0.5d$, and $0\leq z \leq d$ where $d$ is used as our characteristic lengthscale.

We use the same non-dimensionalisation as in \citetalias{duguid_shear_driven_2023} (scaling lengths with respect to $d$, and time with respect to the isothermal sound-crossing time) which results in a system of dimensionless, nonlinear differential equations describing the temporal evolution of the density $\rho$, temperature $T$, velocity $\boldsymbol{u}$ and magnetic field $\boldsymbol{B}$. 
We repeat the equations here for reference, \begin{subequations}\label{maths_p3d_equations}
	\begin{align}
		\rho \dfrac{\partial  \boldsymbol{u}}{\partial t} + \rho (\boldsymbol{u} \cdot \nabla) \boldsymbol{u}&= -{\text{Ta}_0}^{1/2} \, \sigma \kappa \rho \boldsymbol{\Omega}\times \boldsymbol{u} - \nabla P + \theta(m+1) \rho \boldsymbol{e}_z   \nonumber \\
		& \quad \quad + F(\nabla \times \boldsymbol{B})\times \boldsymbol{B} + \sigma \kappa \nabla \cdot \tau + \boldsymbol{F}_s \,,\label{maths_momentum}\\
		\rho\dfrac{\partial T}{\partial t} + \rho (\boldsymbol{u} \cdot \nabla) T & = -(\gamma -1) P \nabla \cdot \boldsymbol{u} + \gamma \kappa \nabla^2 T \nonumber \\
		& \quad \quad + F ( \gamma -1) \zeta_0 \kappa \lvert\nabla\times\boldsymbol{B} \rvert^2 + \frac{(\gamma-1) \sigma \kappa}{2} \tau^2   \,,\label{maths_temperature_equation}\\
		\dfrac{\partial \boldsymbol{B}}{\partial t} & = \nabla \times (\boldsymbol{u} \times \boldsymbol{B} - \zeta_0 \kappa \nabla \times \boldsymbol{B}) \,,\label{maths_induction}\\
		\dfrac{\partial \rho}{\partial t} &= - \nabla \cdot (\rho \boldsymbol{u}) \,,\label{maths_continuity}\\ 
		\nabla \cdot \boldsymbol{B} & = 0 \label{maths_solenoidal_magnetic_field}\,.
	\end{align}
\end{subequations}
Here $\tau$ is the non-dimensional viscous stress tensor
\begin{align}
	\tau_{ij} & \equiv \dfrac{\partial u_i}{\partial x_j} + \dfrac{\partial u_j}{\partial x_i} - \frac{2}{3} \delta_{ij} \dfrac{\partial u_k}{\partial x_k} \label{maths_stress_tensor}
\end{align}
(where $\delta_{ij}$ denotes the Kronecker delta), and the pressure $P$ satisfies the equation of state for an ideal gas, 
\begin{align}
  P=\rho T\,.
\end{align}
The non-dimensional parameters along with their typical values in this work are summarised in Table~\ref{table_dimensionless_parameters}.
Most of these parameters retain the same values as in \citetalias{duguid_shear_driven_2023} to make direct comparisons simpler.
Note that the values for the Taylor number that we quote later are measured at the mid-depth of the domain, $z = 0.5$ (the location of the shear layer), and are related to $\text{Ta}_0$ by $\text{Ta} \approx 55 \text{Ta}_0$.

\begin{table}
	\begin{center}
\def\arraystretch{2}
	\begin{tabular}{ c | c | c | c}
		 & Description & Definition  & Values \\ \hline
		$F$ & Magnetic field strength & $\dfrac{B_0^2}{\mathfrak{R} T_0 \rho_0 \mu_0}$ & $\{0,2.5\times10^{-6}\}$ \\
		$\sigma$ & Prandtl number & $\dfrac{\mu c_p}{K}$ & $0.00025$ -- $0.01$  \\
		$\theta$ & Temperature gradient & $\dfrac{\Delta T}{T_0}$ & 5 \\
		$\kappa$ & Thermal diffusivity & $\dfrac{K}{d \rho_0 c_p \sqrt{\mathfrak{R}T_0}}$ & 0.01	 \\
		$\zeta_0$ & Inverse Roberts number & $\dfrac{\eta c_p \rho_0}{K}$  & ${}^\ast$ $5.0 \times 10^{-4}$  \\
		$\gamma$ & Ratio of specific heats & $\dfrac{c_p}{c_v}$ & $5/3$ \\
		$m$ & Polytropic index & $\dfrac{g d}{\mathfrak{R} \Delta T} -1$  & 1.6 \\
		$\text{Ta}_0$ & Taylor number (top) &  $\dfrac{4\rho_0^2 \Omega^2 d^4}{\mu^2}$  & - \\
  	$\text{Ta}$ & Taylor number (mid)& $\text{Ta}_0\left(1+\dfrac{\theta}{2}\right)^{2m}$   & $10^7$ -- $5\times10^8$ \\
		$\phi$ & Latitude & -  & $\{-\frac{\pi}{2}, -\frac{\pi}{4}, -\frac{\pi}{6}\}$  \\
		$A$		& Shear amplitude & - & 0.02 \\ 
	\end{tabular}
\caption{Non-dimensional parameters in the system, including a text description/name of the quantity, the definition, and the value (or range of values) the parameter takes.
Full details regarding the definitions of these parameters can be found in \citetalias{duguid_shear_driven_2023}. ${}^\ast$ This value for $\zeta_0$ is that adopted for almost all of the presented results, although we comment briefly on the impact of increasing this parameter in \S~\ref{section_diffusion_results}.}
\label{table_dimensionless_parameters}
\end{center}
\end{table}	

All variables are assumed to be periodic in the horizontal directions. 
The upper and lower boundaries are modelled as impermeable and stress-free, with a vanishing tangential magnetic field. 
We also impose the temperature at the upper boundary ($z=0$) and the heat flux at the lower boundary ($z=1$).  
In our non-dimensional units, these boundary conditions have the form
\begin{subequations}\label{maths_bounday_conditions}
\begin{gather}
	u_z = \dfrac{\partial u_x}{\partial z} = \dfrac{\partial u_y}{\partial z} = 0  \quad \text{for} \quad z\in\{0,1\}\,,\\
	B_x = B_y = \dfrac{\partial B_z}{\partial z} = 0 \quad \text{for} \quad z\in\{0,1\}\,,\\
	T = 1  \quad  \text{for} \quad z=0\,,  \quad \text{and}  \quad  \dfrac{\partial T}{\partial z} = \theta  \quad \text{for}  \quad  z=1\,.
\end{gather}
\end{subequations}
The forcing term $\boldsymbol{F}_s$ in Eq.~(\ref{maths_momentum}) is chosen so that, in the absence of magnetic fields or instabilities, it maintains a balance between viscous and Coriolis forces with a ``target'' azimuthal shear flow $U_0(z)\boldsymbol{e}_x$:

\begin{align} \label{maths_shear_forcing}
\boldsymbol{F}_s & \equiv - \sigma \kappa \begin{bmatrix}
           U_0'' \\
           \sqrt{\text{Ta}_0}\, \rho U_0 \sin \phi  \\
           -\sqrt{\text{Ta}_0}\, \rho U_0 \cos \phi
         \end{bmatrix} \,,
\end{align}

\noindent where $'$ denotes a $z$-derivative.
The target flow is the same as in \citetalias{duguid_shear_driven_2023}:
\begin{equation}
    U_0(z) \equiv A \tanh [ 10(z-0.5) ]  \,,
\end{equation}
where $A$ sets the shear amplitude.
For reasons described earlier, and as in \citetalias{duguid_shear_driven_2023}, the flow is allowed to settle to the target profile $U_0$ before we introduce a uniform vertical magnetic field and small-amplitude noise to the temperature.
All simulations presented later adopt the same domain size as in \citetalias{duguid_shear_driven_2023} with $(Lx,Ly,1) = (2,0.5,1)$ and are performed with a numerical resolution $(N_x,N_y,N_z) = (192,128,192)$.
The parameter values were chosen following a more extensive low-resolution parameter survey with $(N_x,N_y,N_z) = (128,64,128)$.

\begin{figure}
	\begin{center}
    \includegraphics[width=\columnwidth]{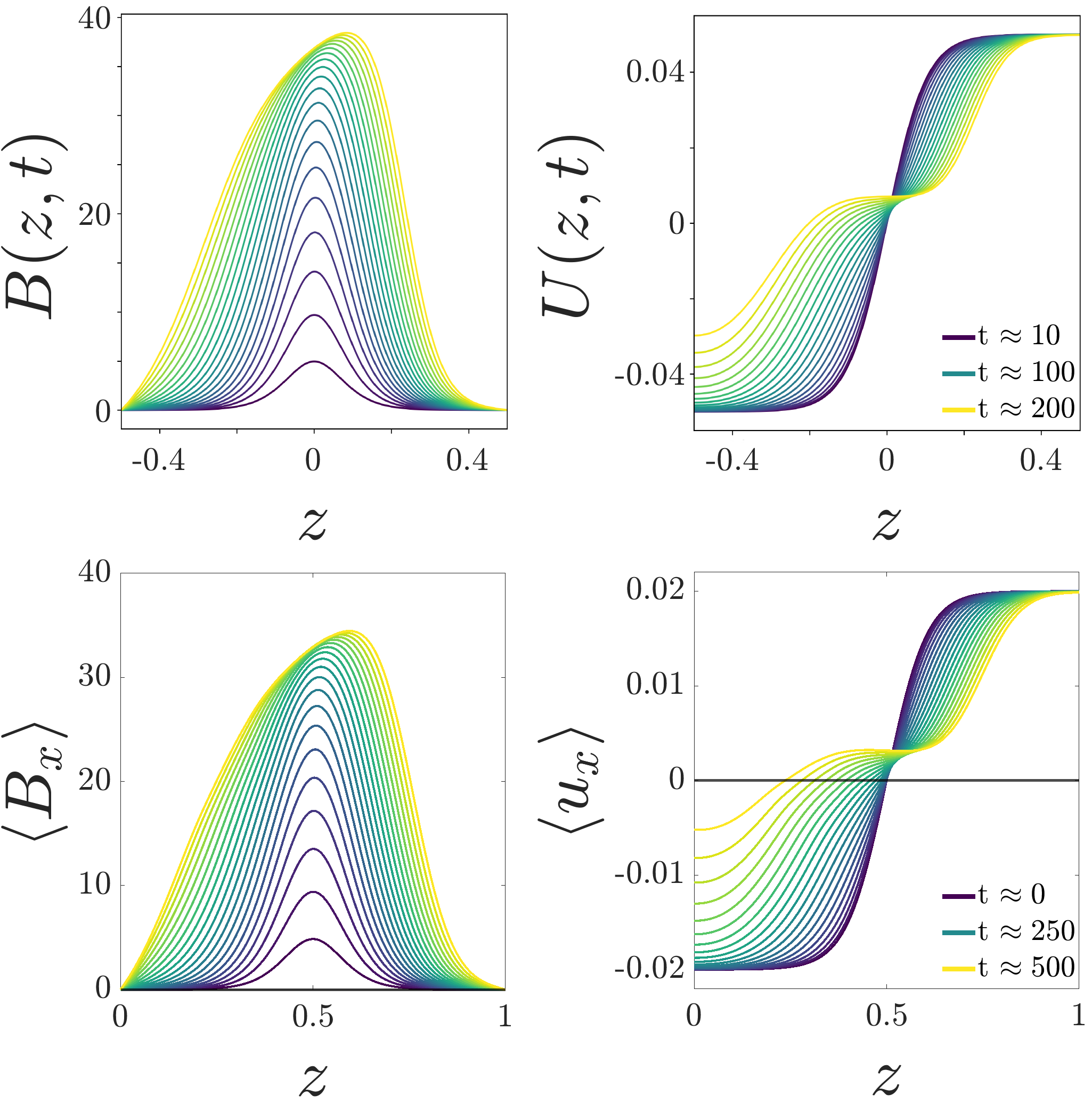}
	\caption{Comparison between a solution of the horizontally-invariant equations (top panels, adapted from \protect\citet{lewis_fluid_2022}) and the full three-dimensional equations (bottom panels, taken from \protect\citet{duguid_shear_driven_2023}, where the angled brackets denote a horizontal average).
    The two cases have different values for the initial vertical field, $F$, and shear amplitude, $A$, leading to dynamics on different timescales, but the profiles of the horizontally-averaged $B_x$ (left panels) and $u_x$ (right panels) are qualitatively comparable.}
	\label{figure_lewiscomparison}
	\end{center}
\end{figure}

In \citetalias{duguid_shear_driven_2023}, we considered the low Prandtl number regime ($\sigma=0.00025$) of \citet{silvers_double-diffusive_2009}, adopting a vertical rotation vector. 
Before proceeding to describe the new results in the present paper, it is useful to first summarise the general evolution of a typical simulation from this low Prandtl number regime. 
In the early stages, the shear flow $U_0(z)\boldsymbol{e}_x$ deforms the initially vertical magnetic field, producing a predominantly horizontal magnetic layer (aligned with the $x$-axis) around $z=0.5$, the peak amplitude of which initially grows linearly in time.
Once the magnetic field becomes strong enough to become dynamically significant, the resultant Lorentz force causes the flow to depart from $U_0$.
This eventually drives the system away from a tachocline-like profile, significantly limiting the field amplification process (the field also diffuses due to resistivity, but this plays only a minor role in the evolution of the large-scale field over the timescale of the simulations).
Due to the significant density stratification, the flow and the field develop vertical asymmetry, with more significant changes occurring in the upper half of the domain.
As shown in Fig.~\ref{figure_lewiscomparison}, the temporal evolution of the magnetic layer is comparable to that of the horizontally-invariant system considered by \citet{lewis_fluid_2022}.
Due to the effects of rotation, the Lorentz force from the growing magnetic layer also drives a persistent flow in the $y$-direction,
as the Lorentz, Coriolis and viscous forces must remain approximately in balance on long timescales.
In cases where the magnetic layer becomes unstable, we observe the formation of tube-like structures in the flow and field (in the upper part of the layer) that are elongated in the direction of the mean field, as expected for magnetic buoyancy instability.
For the cases considered in \citetalias{duguid_shear_driven_2023}, the magnetic buoyancy instability is quasi-two-dimensional in the early stages (before the shear flow departs from the tachocline-like profile), taking the form of an interchange instability, with no significant arching of the magnetic field lines.

\subsection{Mean EMF}
The most relevant quantity for assessing the system's potential to act as a dynamo is the mean electromotive force (mean EMF). 
We compute the mean EMF following the standard mean-field approach \citep[e.g.][]{moffatt_self-exciting_2019}. 
The velocity and the magnetic field are decomposed into their mean and fluctuating parts, $\bullet \equiv \langle \bullet \rangle + \bullet^\prime$. 
Here, angled brackets denote horizontally-averaged quantities defined by
\begin{align}
	\langle \bullet \rangle \equiv \frac{1}{L_x L_y} \int_0^{L_x} \int_0^{L_y} \bullet \, \text{d}x\, \text{d}y\,,
\end{align} 
\noindent where $L_x$ and $L_y$ define the domain size in the horizontal directions (we will make use of angled brackets throughout this paper to denote horizontal averaging of this form), and primes denote fluctuating quantities.
Applying this decomposition to the horizontally-averaged induction equation (Eq.~\ref{maths_induction}), and noting that horizontal averages of fluctuating quantities vanish, we obtain
\begin{align}
	\dfrac{\partial \langle \boldsymbol{B} \rangle}{\partial t} =  \nabla \times \Big(\langle \boldsymbol{u} \rangle \times \langle \boldsymbol{B}\rangle + \mathcal{E}\Big) + \zeta_0 \kappa \dfrac{\partial^2}{\partial z^2} \langle \boldsymbol{B} \rangle\,,
 \label{eq:mean-field}
\end{align}
where the mean EMF, $\mathcal{E}$, is defined as
\begin{align}
	\mathcal{E} \equiv \langle \boldsymbol{u}^\prime \times \boldsymbol{B}^\prime \rangle\,.
\end{align}
We note that only the $x$ and $y$ components of $\mathcal{E}$ contribute to the generation of the mean field in equation~(\ref{eq:mean-field}).
In idealised mean-field theory, the mean EMF is often expressed as a sum of contributions from the $\alpha$~effect, turbulent pumping, and turbulent magnetic diffusion.
However, as noted by \citet{davies_mean_2011}, and discussed in detail in \citetalias{duguid_shear_driven_2023}, it is not clear that such a decomposition is meaningful in the present context.
In what follows, we will focus upon the EMF itself rather than its interpretation in terms of mean-field theory.

\section{Varying the viscous and magnetic diffusivities}\label{section_diffusion_results}

For plausible velocity shear profiles, shear-driven magnetic buoyancy has only previously been achieved in numerical simulations by adopting values for the viscosity and magnetic diffusivity that are much smaller than the thermal diffusivity. 
In particular, the simulations presented in \citetalias{duguid_shear_driven_2023} had $\zeta_0 = 5\times10^{-4}$ and $\sigma = 2.5\times10^{-4}$ (with a dimensionless thermal diffusivity of $\kappa = 0.01$). 
Although these simulations are far more dissipative than the real tachocline, these parameters do ensure that the viscous, Ohmic and thermal diffusion timescales have the correct ordering, with the thermal diffusion time much shorter than the other two.
However, as noted in the Introduction, this is a challenging numerical regime.
Furthermore, the simulations described in \citetalias{duguid_shear_driven_2023} had a number of other limitations, particularly the eventual deviation of the mean velocity from the target profile.
In this section we will investigate to what extent these results depend on the values of $\sigma$ and $\zeta_0$, and whether the constraints on their values can be relaxed. 
As well as reducing the computational burden we would ideally also like to find a parameter regime in which the shear flow remains tachocline-like, with strong azimuthal shear (in the $x$ direction) and much weaker flows in the latitudinal ($y$) direction.

In order to make direct comparisons with previous results we will retain most of the parameters from \citetalias{duguid_shear_driven_2023} (see Table~\ref{table_dimensionless_parameters}).
Throughout this section we take the rotation axis to be vertical, with $\phi = -\pi/2$ (corresponding to the south pole)\footnote{This is the same rotation vector considered in \citetalias{duguid_shear_driven_2023}, but note that there is an erroneous minus sign in the definition of $\boldsymbol{\Omega}$ in that paper, which should read $\boldsymbol{\Omega} = \Omega\mathbf{e}_z$ rather than $\boldsymbol{\Omega} = -\Omega\mathbf{e}_z$.} and the rotation rate is fixed, with a mid-layer Taylor number of $\text{Ta} = 5\times 10^8$, matching the most rapidly rotating case in \citetalias{duguid_shear_driven_2023}.
With the (dimensionless) thermal diffusivity fixed at $\kappa=0.01$ (as in \citetalias{duguid_shear_driven_2023}), we continue to restrict attention to values of $\sigma$ and $\zeta_0$ that are much smaller than unity, to ensure that the thermal diffusion timescale is always shorter than the other diffusive timescales.
However, there is certainly scope to increase both $\sigma$ and $\zeta_0$ without violating this condition. 
Whilst this does push the calculations slightly further away from the conditions in the tachocline, the advantage of adopting this approach is that it makes the computations less onerous (which will be essential for future simulations of the full dynamo problem). 

We focus initially upon the effects of increasing $\zeta_0$, i.e.~increasing the magnetic diffusivity.
Unfortunately, even a modest increase from $\zeta_0 = 5.0 \times 10^{-4}$ to $\zeta_0 = 1.5 \times 10^{-3}$ suppresses the instability in our simulations.
This result is partly explained by enhanced diffusion of the magnetic layer, which reduces the strength and gradient of $\langle B_x \rangle$.
Much more importantly, however, the increased value of $\zeta_0$ increases the magnetic field gradient required for magnetic buoyancy instability to occur \citep{gilman_instability_1970,acheson_instability_1979,hughes_tachocline_2007}.
Based on this finding, and given that we do not want to decrease $\zeta_0$ any further for numerical reasons, we therefore fix $\zeta_0 = 5 \times 10^{-4}$ in all calculations that will be presented from here on.
The remainder of this section will focus upon the (much more interesting) changes in behaviour that occur when the Prandtl number is varied. 

\subsection{Prandtl number}\label{section_Pr}

\begin{figure*}
\begin{center}
	\includegraphics[width=\textwidth]{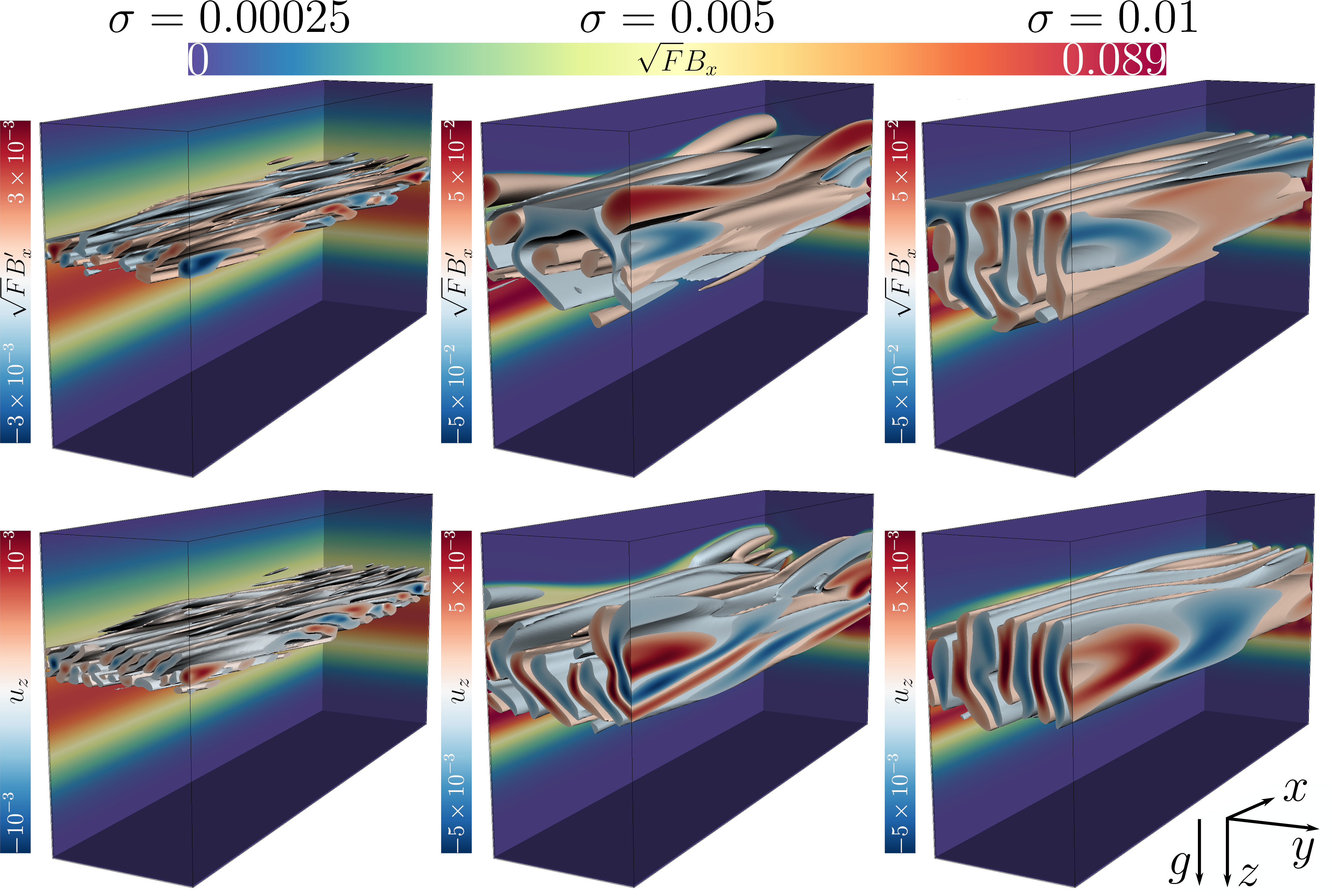}
	\caption{Snapshots taken at $t\approx 500$ for $\sigma \in \{0.00025, 0.005, 0.01\}$.  Top: pseudocoloured isovolumes of $\sqrt{F}B_x^\prime$, where $B_x^\prime = B_x - \langle B_x \rangle$, with $\sqrt{F}\langle B_x \rangle$ as a
    backdrop. Bottom: pseudocoloured isovolumes of $u_z$ again with $\sqrt{F}\langle B_x \rangle$ as a backdrop.  
}
	\label{figure_diff_snapshots}
\end{center}
\end{figure*}

Guided by an extensive low-resolution parameter survey, we will present results for $\sigma \in \{0.00025, 0.005, 0.01\}$. 
We demonstrated in \citetalias{duguid_shear_driven_2023} that the $\sigma = 0.00025$ case does indeed generate magnetic buoyancy instability; it is provided here as a reference point for the other cases. 
At the other end of this range, we elected not to increase $\sigma$ beyond $0.01$ for a number of reasons. 
In particular, we wanted to ensure that the effects of viscous heating (associated with the forced shear) remained negligible on the timescales of interest, and it has been confirmed (by artificially suppressing the relevant terms in the code) that this is indeed the case even for $\sigma=0.01$. 
Secondly, we needed to ensure that the thermal diffusion timescale remained much shorter than the viscous timescale, and increasing $\sigma$ much further would have violated this constraint.
Finally, for larger values of $\sigma$ we found that the initial perturbations dissipated very quickly before the magnetic field became buoyantly unstable, significantly delaying the apparent onset of the instability (for a given initial thermal perturbation).
Given that the magnetic layer is itself evolving in time, delayed onset for the magnetic buoyancy instability could have detrimental consequences for any potential future dynamo calculation. 

There is one further consideration that should be mentioned before presenting the results, namely the stability of the background shear.
As discussed in \citetalias{duguid_shear_driven_2023}, shear flows of this nature can be hydrodynamically unstable if the product of the Prandtl number and the Richardson number, $\text{Ri}$ (which is the square of the Brunt-V\"ais\"al\"a frequency divided by the local shearing rate), is small \citep{ledoux_rotational_1974,garaud_turbulent_2017}.
For the simulations described here, for which $\text{Ri} \approx 18.57$, we have $\text{Ri}\,\sigma \approx \{0.0046, 0.0929, 0.1857\}$ for $\sigma \in \{0.00025, 0.005, 0.01\}$ respectively. 
The two higher $\sigma$ cases lie well above the approximate stability threshold of $\text{Ri}\,\sigma > 0.007$ determined empirically by \citet{garaud_turbulent_2017}, which suggests that both should be stable. 
This has been confirmed by a set of hydrodynamic simulations. 
In the absence of a magnetic field, any deviations from the target flow (which might be indicative of an underlying hydrodynamic instability) are even weaker in these cases than that shown for the lowest $\sigma$ case in \citetalias{duguid_shear_driven_2023}. 

Figure~\ref{figure_diff_snapshots} illustrates the key differences in the flow and field structures as a function of $\sigma$ after $t\approx500$ time units. 
The upper panels show the distributions of $\langle B_x \rangle$ and $B_x^\prime$; the lower panels show the distribution of $u_z$ (note that the $B_x$ plots include a factor of $\sqrt{F}$, which allows a more direct comparison with the amplitude of $u_z$ in our dimensionless units).
These snapshots show that the magnetic buoyancy instability for each value of $\sigma$ is at a different stage of its development.
For  $\sigma = 0.00025$, the instability is confined to a thin layer and the perturbations have a relatively low amplitude, with very short lengthscales in both of the field-perpendicular ($y$ and $z$) directions.
As was highlighted in \citetalias{duguid_shear_driven_2023}, the instability for this case is interchange-like, in the sense that the perturbations have little variation in the direction parallel to the mean field.
The perturbations in both of the higher $\sigma$ cases have reached a significantly greater amplitude at $t\approx 500$, because in those cases the instability has both an earlier onset and a faster growth rate.
As mentioned earlier, in those cases the instability is initially interchange-like, but by the time shown in Fig.~\ref{figure_diff_snapshots} the perturbations have developed a clear undular structure.
The fluid motions by this time have developed into plumes, with a ``mushroom''-like cross-section \citep[reminiscent of that observed by][]{cattaneo_nonlinear_1988}.
As would be expected, this undular structure is characterised by a long lengthscale in the field-parallel direction, and a much shorter lengthscale in the $y$ direction, although it is apparent that the typical lengthscale in the $y$ direction tends to increase with increasing $\sigma$. 
One other difference that is worth mentioning at this stage is the relative tilt of the tube-like structures in the $xy$-plane (a phenomenon that was discussed in detail in \citetalias{duguid_shear_driven_2023}).
Although it is not easy to see in these figures, there is quite a pronounced tilt in the lowest $\sigma$ case; the degree of tilt for larger $\sigma$ is much smaller.

\begin{figure}
	\begin{center}
		\includegraphics[width=\columnwidth]{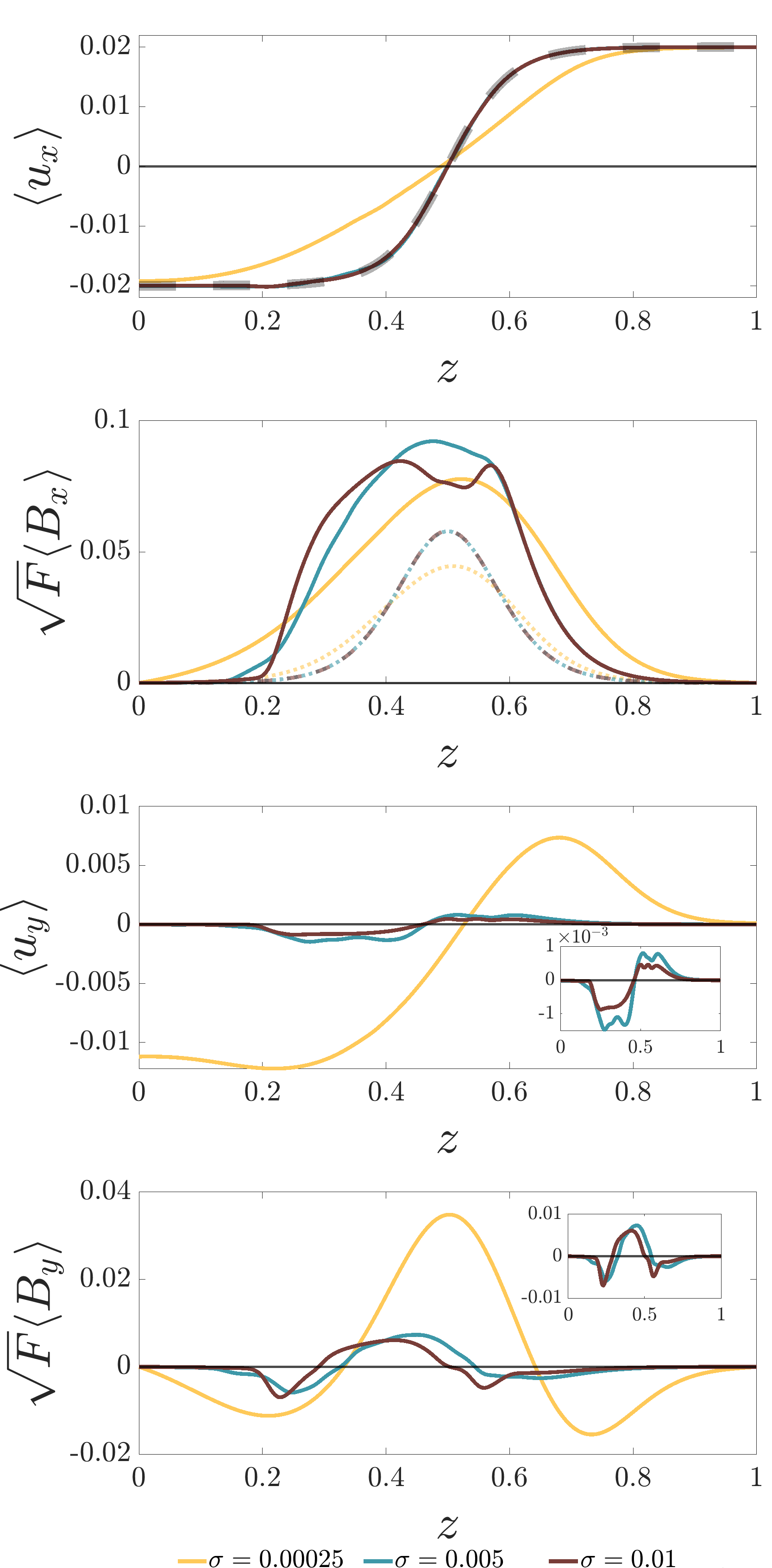}
		\caption{Horizontally-averaged vertical profiles taken at $t\approx 500$ for $\sigma\in \{0.00025, 0.005, 0.01\}$, as denoted by the legend. 
        Note that for $\sqrt{F}B_x$ we also show the profiles at $t\approx 200$ (denoted by dotted/dashed lines) to highlight the evolution of the field. 
        The insets in the lower two plots focus upon the higher $\sigma$ cases.}
		\label{figure_diff_uxBxuyBy}
	\end{center}
\end{figure}

The Prandtl number dependence that is illustrated by Fig.~\ref{figure_diff_snapshots} can be understood by analysing the underlying mean fields and flows.
The top panel of Fig.~\ref{figure_diff_uxBxuyBy} shows the horizontally-averaged vertical profiles of $u_x$ at $t\approx 500$ for each $\sigma$.
As noted in \citetalias{duguid_shear_driven_2023} for the lowest $\sigma$ case, the velocity shear is reduced by the action of the Lorentz force, eventually departing significantly from the target flow, $U_0$.
In the higher $\sigma$ cases, by contrast, the shear flow is practically indistinguishable from $U_0$.
Given the importance of the shear profile for the generation of the magnetic layer, it is not surprising that we see significant differences in the horizontally averaged vertical profiles of $B_x$, which are also shown in Fig.~\ref{figure_diff_uxBxuyBy}.
At early times, in all cases, the induced field is strongest about the mid-plane.
At $t\approx 200$, we see that the $\langle B_x\rangle$ distributions are almost identical in the higher $\sigma$ cases, whereas the lowest $\sigma$ case is already diverging as the flow is being perturbed by the magnetic field (producing a shallower magnetic field gradient above the mid-plane). 
Whilst the peak magnetic fields are of the same order of magnitude in all cases at $t\approx 200$, the Prandtl number dependence of the forcing (Eq.~\ref{maths_shear_forcing}) means that the field can more easily perturb the force balance in the lowest $\sigma$ cases, leading to the observed flattening of the shear profile. 
At later times, we see a smoother $\langle B_x\rangle$ profile in the $\sigma = 0.00025$ case, whilst there is much more structure in the mean field distribution in the higher $\sigma$ cases.
This is a signature of the vigour of the magnetic buoyancy instability at higher Prandtl number, which is strong enough to significantly alter the vertical distribution of the horizontal magnetic field. 
The relative vigour of the instability in these cases is simply a consequence of a steeper vertical gradient in the shear-generated magnetic layer, which tends to promote instability \citep{gilman_instability_1970,acheson_instability_1979,hughes_tachocline_2007}.

\begin{figure}
	\begin{center}
    \includegraphics[width=\columnwidth]{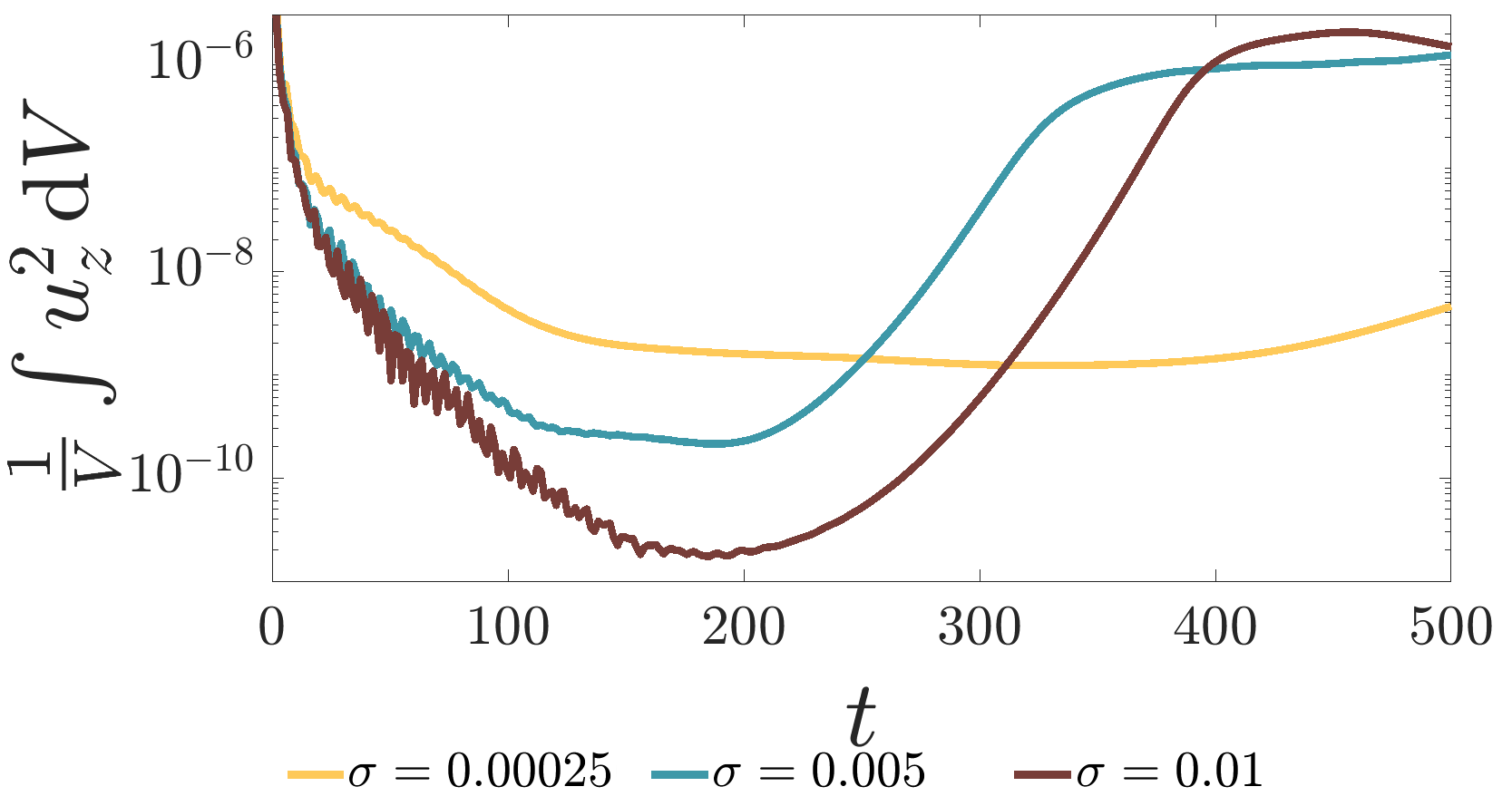}
	\caption{Time series of the volume averaged vertical velocity energy for each value of $\sigma$ (see legend).}
	\label{figure_uzprime_timeseries_diff}
	\end{center}
\end{figure}

As discussed in \citetalias{duguid_shear_driven_2023}, the inclusion of rotation tends to drive mean flows in the $y$-direction once the magnetic field becomes large enough to disturb the initial hydrodynamic force balance (in which the imposed body force balances viscosity and the Coriolis force that results from the driven shear). 
For the reasons outlined above we might again expect to see weaker flows in this field-perpendicular direction at higher $\sigma$. 
This can be seen to be the case in the third plot of Fig.~\ref{figure_diff_uxBxuyBy}, which shows the horizontally-averaged vertical profiles of $u_y$ for each Prandtl number at $t\approx 500$.
In the $\sigma = 0.00025$ case, these flows reach a peak amplitude that is comparable in magnitude to that of the initial shear.
Whilst there is also a systematic $\langle u_y\rangle$ flow in the higher $\sigma$ cases, its amplitude is negligible (more than an order of magnitude smaller) compared to the mean flow in the $x$-direction.
Since $\langle B_y\rangle$ is being induced by $\langle u_y \rangle$, at a rate approximately proportional to $\partial_z \langle u_y \rangle$, we observe a decrease in the magnitude of $\langle B_y \rangle$ for increasing $\sigma$ (in the lower part of Fig.~\ref{figure_diff_uxBxuyBy}).
To put this another way, whilst the mean horizontal field exhibits a significant tilt in the $xy$-plane in the $\sigma = 0.00025$ case (which significantly complicates the interpretation of the mean EMF measurements, as discussed in \citetalias{duguid_shear_driven_2023}), this tilting effect tends to be reduced as we increase $\sigma$.
This lack of tilt is reflected in the field and flow structures that are shown in Fig.~\ref{figure_diff_snapshots}.

To analyse the onset (and subsequent evolution) of the magnetic buoyancy instability in a more quantitative manner, Fig.~\ref{figure_uzprime_timeseries_diff} shows the volume-averaged energy in the vertical velocity for $\sigma \in \{0.00025, 0.005, 0.01\}$. 
All cases feature an initial decay phase as the initial, thermally-induced perturbations gradually dissipate. 
Note that the oscillatory behaviour at early times is indicative of the presence of acoustic and internal gravity waves, the latter of which are dominant. 
In both the $\sigma = 0.00025$ and $\sigma=0.005$ cases, the perturbation energy plateaus at $t\approx 150$ (at a lower level in the latter case).
For $\sigma=0.005$, the magnetic buoyancy instability onsets at around $t\approx 200$, and we observe a rapid growth phase until saturation occurs at around $t \approx 300$. 
In the lowest $\sigma$ case, the instability is clearly delayed (due to the smoothing of the shear profile), and only at around $t\approx 400$ do we start to see well-defined indications of growth. 
In the $\sigma=0.01$ case, the initial decrease in the energy in the vertical velocity plateaus slightly later, and at a much lower level (due to the enhanced diffusion).
However the magnetic buoyancy instability again sets in at $t\approx 200$, with the energy in the vertical velocity then growing at a similar rate to the $\sigma=0.005$ case. 
Here the instability saturates slightly later, at $t\approx 400$, because it takes longer for the amplitude of the perturbations to grow to the required level for saturation.   
Further increases in $\sigma$ would further reduce the amplitude of the remaining perturbation at the critical time for the onset of the magnetic buoyancy instability (at $t \approx 200$), further delaying the point at which the instability starts to produce dynamically-significant perturbations to the field and flow. 

\subsection{Mean electromotive force}

\begin{figure*}
\begin{center}
	\includegraphics[width=\textwidth]{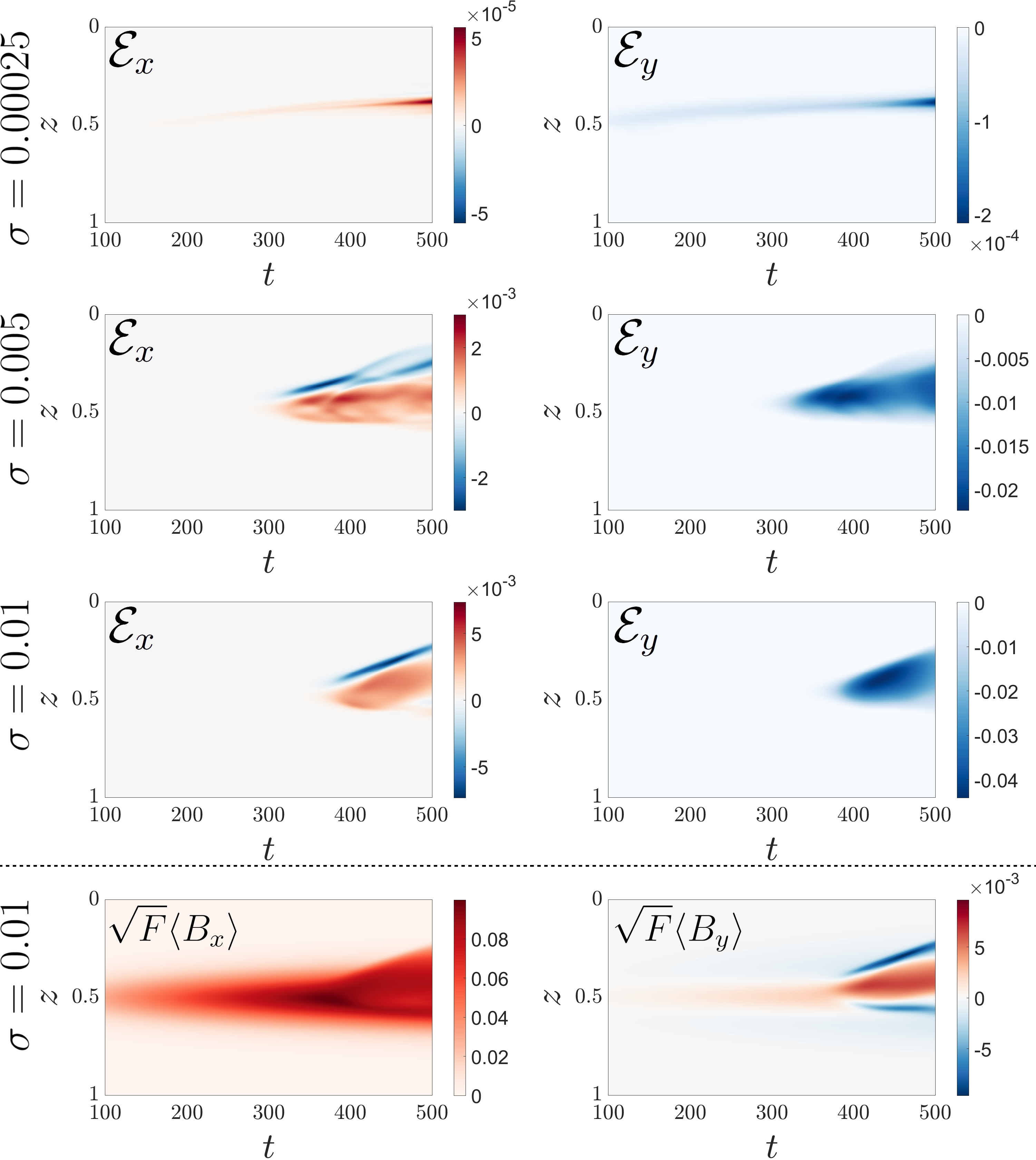}
	\caption{The depth- and time-dependence of $\mathcal{E}_x$ and $\mathcal{E}_y$ for $\sigma\in \{0.00025, 0.005, 0.01\}$ (upper three rows). In the bottom row, we also show (for the purposes of comparison) the depth- and time-dependence of $\sqrt{F}\langle B_x \rangle$ and $\sqrt{F}\langle B_x \rangle$ for $\sigma = 0.01$.}
	\label{figure_EMF_diff}
\end{center}
\end{figure*}

With a detailed understanding of the behaviour of the flow and field as $\sigma$ is varied, we now look at how this influences the mean EMF components, $\mathcal{E}_x= \langle u_y^\prime B_z^\prime -u_z^\prime B_y^\prime\rangle$ and $\mathcal{E}_y=\langle u_z^\prime B_x^\prime -u_x^\prime B_z^\prime\rangle$.
As noted above, these are the only components of the mean EMF that contribute to the generation of the mean magnetic field in Eq.~(\ref{eq:mean-field}).
Although there are some difficulties in terms of the standard mean-field interpretation of the mean EMF in the context of this magnetically-driven instability, insights from mean-field theory suggest that the component of the mean EMF that is parallel to the mean magnetic field should play a crucial regenerative role in any Parker-like dynamo mechanism.

The lowest Prandtl number case ($\sigma = 0.00025$) was discussed in detail in \citetalias{duguid_shear_driven_2023}, so we only briefly summarise the main results here.
The time- and depth-dependence of the $x$ and $y$ components of the mean EMF for $\sigma = 0.00025$ are shown in the upper panels of Fig.~\ref{figure_EMF_diff}. 
In this case $\mathcal{E}_x$ has a systematic positive-signed localised peak that grows super-linearly in time.
The peak first appears near the mid-plane and then migrates upwards, roughly following the evolution of the buoyancy instability.
There is also a substantial negative $\mathcal{E}_y$, whose location closely follows that of $\mathcal{E}_x$.
Furthermore, the magnitude of $\mathcal{E}_y$ significantly exceeds that of $\mathcal{E}_x$.
\citet{davies_mean_2011}, who observed similar behaviour in their idealised linear system, associate the component of the EMF that is perpendicular to the mean field with magnetic pumping and turbulent diffusion.  
Although the magnitude of the EMF is small (for example, $\mathcal{E}_x$ is approximately $5\times 10^{-5}$ at $t\approx 500$), both components are still growing at the end of the simulation, and might eventually become large enough to influence the evolution of the mean magnetic field.
However, by this time the mean flow has already departed significantly from the imposed ``tachocline-like'' flow, so it is not clear that longer integrations would be meaningful.
The substantial (and depth-dependent) rotationally induced tilt of the mean horizontal field also makes it difficult to disentangle the field-parallel and field-perpendicular contributions to the mean EMF.
This further complicates the interpretation of these measurements.
As noted in the previous section, the rotationally induced tilt of the mean magnetic field is significantly reduced at higher Prandtl numbers, meaning that the geometrical distinction between the field-parallel and field-perpendicular directions is much clearer in these cases.

The second and third rows of Fig.~\ref{figure_EMF_diff} show the time- and depth-evolution of $\mathcal{E}_x$ and $\mathcal{E}_y$ for the higher Prandtl number cases, $\sigma \in \{0.005, 0.01\}$.
For these cases we find  $|\mathcal{E}_x|$ to be approximately two orders of magnitude larger than in the $\sigma = 0.00025$ case at comparable times (at least up to $t\approx 500$).
This is a consequence of a much more vigorous magnetic buoyancy instability (which is itself a consequence of the fact that the shear remains closer to the target shear), which produces stronger perturbations to the field and flow. 
A mean EMF component of this magnitude is much more likely to play a significant role in the evolution of the mean field than the corresponding $\mathcal{E}_x$ component in the $\sigma = 0.00025$ case.
Compared to the lowest Prandtl number case, there is also a more complicated depth dependence for $\mathcal{E}_x$.
Whilst there is still a systematic positive $\mathcal{E}_x$ in the vicinity of the mid-plane, both the $\sigma=0.005$ and $\sigma=0.01$ cases show a very thin negative band that migrates towards the surface as the simulation progresses, following the evolution of the magnetic buoyancy instability.  
For $\sigma \in \{0.005, 0.01\}$ we can see from  Fig.~\ref{figure_EMF_diff} that $\mathcal{E}_y$ remains predominantly negative and is consistently of larger magnitude than $\mathcal{E}_x$.
Furthermore, just like $\mathcal{E}_x$, the region of significant $\mathcal{E}_y$ expands into the upper regions of the domain as time evolves. 
A substantial $\mathcal{E}_y$ therefore seems to be a robust feature of this system, independent of the choice of $\sigma$. 
The bottom row of Fig.~\ref{figure_EMF_diff} shows the time- and depth-evolution of the mean-field components $\sqrt{F}\langle B_x \rangle$ and $\sqrt{F}\langle B_y \rangle$ for the case with $\sigma = 0.01$. 
Although there is a clear correlation between the location of the mean field and that of the mean EMF,
the two are not related in a straightforward (e.g.~linear) fashion.

In order to understand the observed features of the mean EMF, it is instructive to analyse its constituent parts in more detail.
In the case of $\mathcal{E}_x= \langle u_y^\prime B_z^\prime -u_z^\prime B_y^\prime\rangle$, for example, separate consideration of $\langle u_y^\prime B_z^\prime \rangle$ and $-\langle u_z^\prime B_y^\prime \rangle$ yields some important insights (as shown in \citetalias{duguid_shear_driven_2023} for $\sigma=0.00025$). 
However, we can gain further insights by Fourier decomposing the magnetic field and flow perturbations into a sum of components with different wavenumbers in the azimuthal ($x$-)direction. 
The reason for considering such a decomposition is to determine the relative importance of undular and interchange modes in the production of the mean EMF.
Due to the periodicity of the domain, the (dimensionless) azimuthal wavenumbers of these Fourier components take the form $n_x\pi$, where $n_x$ is a non-negative integer representing the number of wavelengths in the $x$-direction. 
Therefore $n_x=0$ corresponds to an interchange mode, whilst $n_x=1$ corresponds to the undular mode with the longest permitted wavelength in the domain, $0 \le x \le 2$.
The mean EMF, and its constituent parts, can then be expressed as sums over the contributions from different values of $n_x$.
We will focus our Fourier analysis on the higher Prandtl number cases below.  
The same analysis can be carried out for the $\sigma=0.00025$ case but the results are qualitatively different.
This is a consequence of the rotationally-induced tilt of the mean magnetic field (which, as noted above, significantly complicates the interpretation of the analysis).

\begin{figure*}
\begin{center}
	\includegraphics[width=\textwidth]{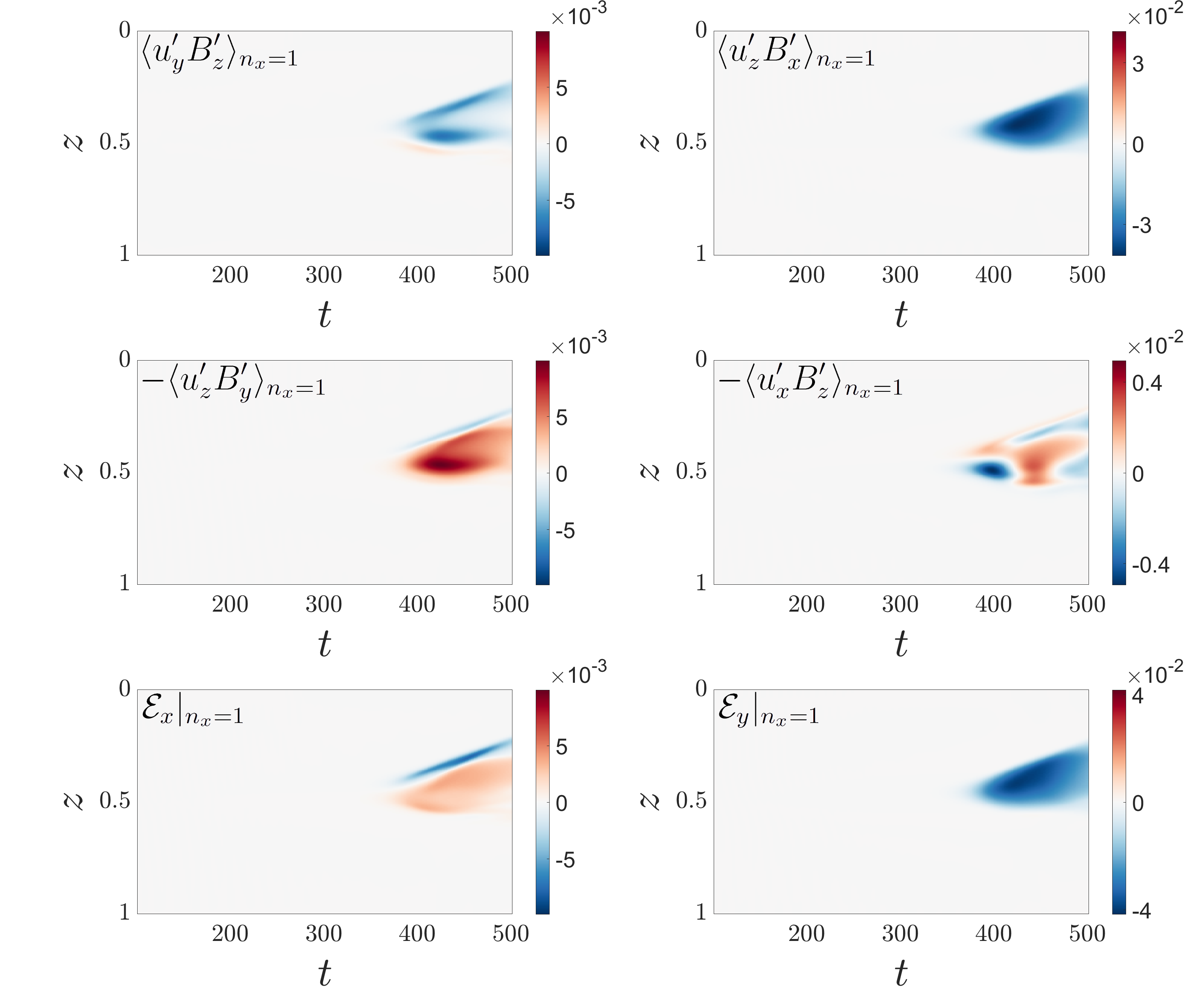}
	\caption{The depth- and time-dependence of $\mathcal{E}_x=\langle u_y^\prime B_z^\prime -u_z^\prime B_y^\prime\rangle$ and $\mathcal{E}_y=\langle u_z^\prime B_x^\prime -u_x^\prime B_z^\prime\rangle$ and the constituent parts of these components, for $\sigma = 0.01$, filtered to show only contributions from flow and field components with the largest lengthscale undular mode $n_x=1$. }
	\label{figure_EMF_diff_kx1}
\end{center}
\end{figure*}

In the larger Prandtl number cases, the dominant contributions to $\mathcal{E}_x$ come from perturbations with $n_x \in \{0,1\}$, which represent the longest wavelengths in the field-parallel direction.
In Fig.~\ref{figure_EMF_diff_kx1} we show the contribution from perturbations with $n_x=1$
in the $\sigma = 0.01$ case.\footnote{In the interests of clarity, the subscript ``$n_x=1$'' here means that only the Fourier modes of $\mathbf{u}'$ and $\mathbf{B}'$ with $n_x=1$ are included.} 
It is clear from comparing Fig.~\ref{figure_EMF_diff} and \ref{figure_EMF_diff_kx1} that the $n_x=1$ mode has the same spatio-temporal structure as the total $\mathcal{E}_x$, as well as a similar magnitude.
The $n_x=0$ mode for this simulation (not shown) is somewhat smaller in magnitude, and has a less coherent structure.
As a result, the $n_x=1$ mode is the dominant component, with the interchange mode (i.e.~$n_x=0$) accounting for most of the small differences between the contribution from $n_x=1$ mode and the total $\mathcal{E}_x$.
Both terms $\langle u_y^\prime B_z^\prime \rangle$ and $-\langle u_z^\prime B_y^\prime \rangle$ of $\mathcal{E}_x$ are of similar magnitude for $n_x \in \{0,1\}$.
The former of these terms is predominantly negative, whereas the latter is predominantly positive, apart from a thin negative band that propagates upwards.
As a result these two terms generally work in opposition, but reinforce one another in the upper band leading to the observed strong negative $\mathcal{E}_x$.
We observe similar features when decomposing the EMF in the case $\sigma = 0.005$ (not shown).
As for $\mathcal{E}_x$, the dominant contributions to $\mathcal{E}_y$ come from the $n_x \in \{0,1\}$ modes of the flow and field.
The $n_x=1$ contribution, which is illustrated in Fig.~\ref{figure_EMF_diff_kx1}, is again the larger of the two.
Further, it is evident that the $u_z^\prime B_x^\prime$ term is the dominant component. 
This analysis clearly shows that the undular ($n_x=1$) magnetic buoyancy mode is playing a crucial role in the production of a substantial mean EMF in the higher Prandtl number cases.

\section{Varying the rotation}\label{section_rotation}

Based on insights from mean-field theory, a lack of reflectional symmetry should be conducive to the production of a systematic mean EMF. 
We might therefore expect to see a strong dependence of these results upon the choice of rotation rate and the inclination of the rotation vector, $\phi$ (which corresponds to the latitudinal location of the local Cartesian domain). 
Given the latitudinal distribution of sunspots, which are never observed near the poles, it is particularly important to consider the effects of varying the inclination of the rotation vector, which we will describe later in this section.
However, we shall first briefly discuss the effects of varying the rotation rate in the case of a vertical rotation vector ($\phi=-\pi/2$).

\subsection{Taylor number dependence}\label{subsection_taylor}

The rotation rate dependence in the lowest Prandtl number case ($\sigma=0.00025$) was clearly demonstrated in \citetalias{duguid_shear_driven_2023}. 
At moderate Taylor numbers, this system produced a coherent $\mathcal{E}_y$ but a negligible $\mathcal{E}_x$.
To produce a more substantial $\mathcal{E}_x$, it was necessary to increase the Taylor number to $\text{Ta}=5\times 10^8$ (which motivated this choice of parameter in the present paper). We have already seen that the higher Prandtl number cases produce a much larger mean EMF for $\text{Ta}=5\times 10^8$ than the case considered in \citetalias{duguid_shear_driven_2023}. 
Furthermore, the increased Prandtl number has a profound impact upon the maintenance and generation of the mean flows, with a more closely-maintained target shear and much weaker mean flows in $y$-direction.
Given these differences, it is not immediately obvious how these higher Prandtl number cases will depend upon $\text{Ta}$.

Taking the case $\sigma = 0.01$, we have reduced the Taylor number from $\text{Ta} = 5\times10^8$ to $10^7$, which we previously found was sufficient to produce qualitative changes
to the dynamics
in the case $\sigma=0.00025$.
For $\sigma=0.01$, by contrast, this reduction yields qualitatively similar dynamics, only on a longer timescale.
In particular, both components of the mean EMF are similar to those in the lower panels of Fig.~\ref{figure_EMF_diff}, except that a longer time is taken to reach comparable magnitudes.
We conclude that, for this larger value of $\sigma$, the dynamics are affected by rotation at a lower value of $\text{Ta}$.
This may be explained by the fact that, in our dimensionless units, increasing $\sigma$ while fixing $\text{Ta}$ corresponds to increasing the rotation rate (which is why the Coriolis term in Eq.~(\ref{maths_momentum}) includes a factor of $\sigma$).

\subsection{Varying the inclination}\label{subsection_incline}

Having noted that the higher Prandtl number cases are not strongly dependent upon the choice of $\text{Ta}$, we now fix $\text{Ta}=5\times 10^8$
(for ease of comparability with previous results) and vary $\phi$. 
In addition to the results presented earlier for $\phi = -\pi/2$ (high latitude) we have therefore also performed calculations for $\phi=-\pi/4$ (mid latitude) and $\phi=-\pi/6$ (low latitude). 
The rest of the parameters are fixed to the values presented in Table~\ref{table_dimensionless_parameters}, focusing exclusively upon the $\sigma=0.01$ case.

\begin{figure*}
\begin{center}
	\includegraphics[width=\textwidth]{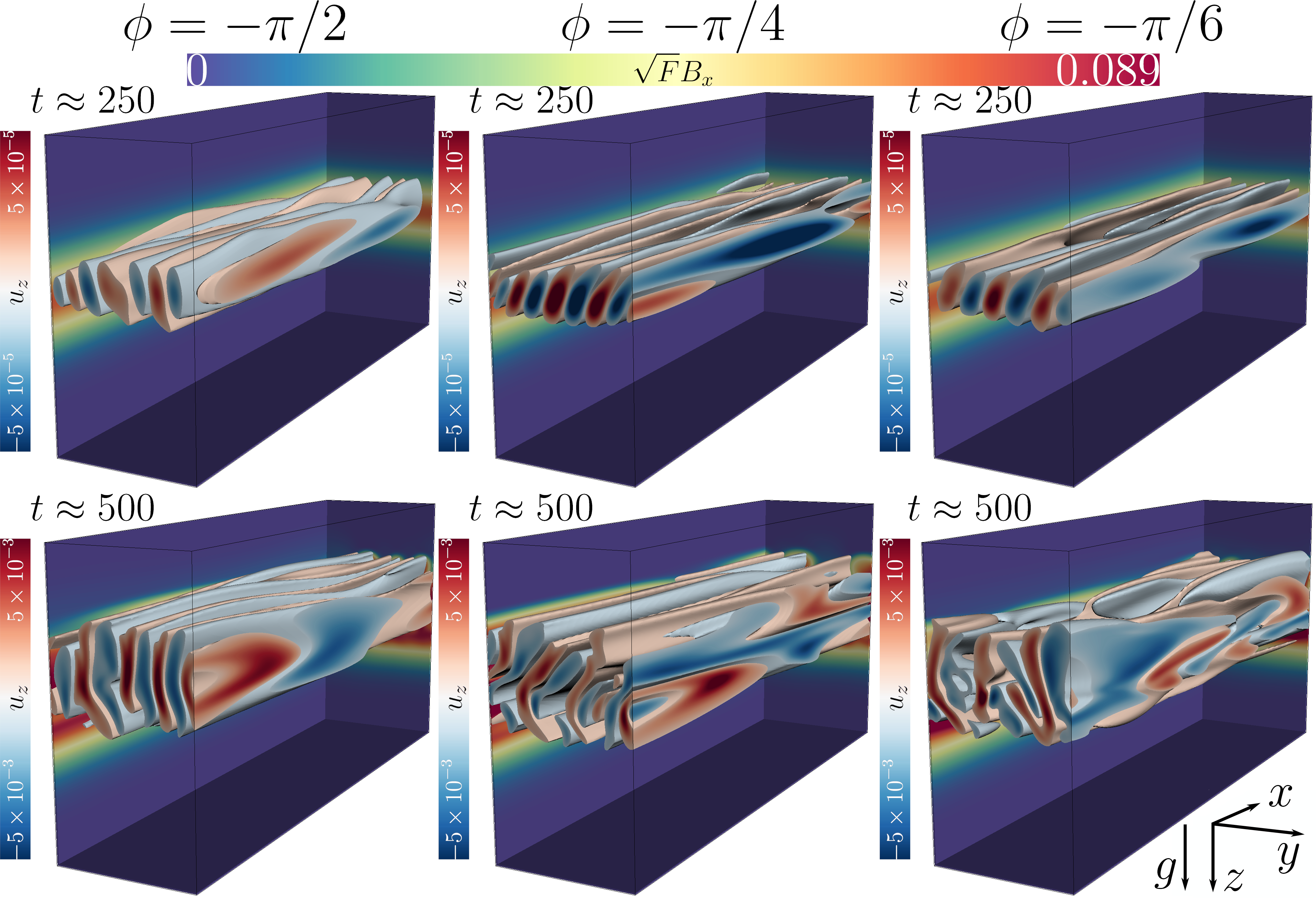}
	\caption{The dependence of the $\sigma=0.01$ case on the inclination of the rotation vector, showing snapshots (left to right) for $\phi\in \{-\pi/2,-\pi/4, -\pi/6\}$. 
    Top: pseudocoloured isovolumes of $u_z$ at $t\approx 250$, with the horizontally-averaged $\sqrt{F} B_x$ as the backdrop.
    Bottom: The same quantities at $t\approx 500$.
}
	\label{figure_tilted_snapshots}
\end{center}
\end{figure*}

Fig.~\ref{figure_tilted_snapshots} shows volume renderings of $u_z$ at $t\approx250$ and $t\approx500$ for each value of $\phi$.
During the early, essentially linear development of the instability ($t\approx250$),
we see that the flow structures are elongated in a direction roughly parallel to the rotation axis, in accordance with the Taylor--Proudman effect.
In all cases the perturbations grow at a similar rate, and nonlinear saturation of the instability occurs at $t\approx400$.
The bottom part of Fig.~\ref{figure_tilted_snapshots} shows a later stage of each of the simulations (at $t\approx500$).
By this point in the evolution, the effects of the tilt have become significantly less apparent. 
In all cases, we see a comparatively vigorous instability, with some indication of undulation
in the field-wise direction (particularly in the $\phi=-\pi/6$ case).
Once the magnetic buoyancy instability has saturated,
the tilting of the rotation vector appears to have little effect on the key features of the dynamics.

\begin{figure}
	\begin{center}
		\includegraphics[width=\columnwidth]{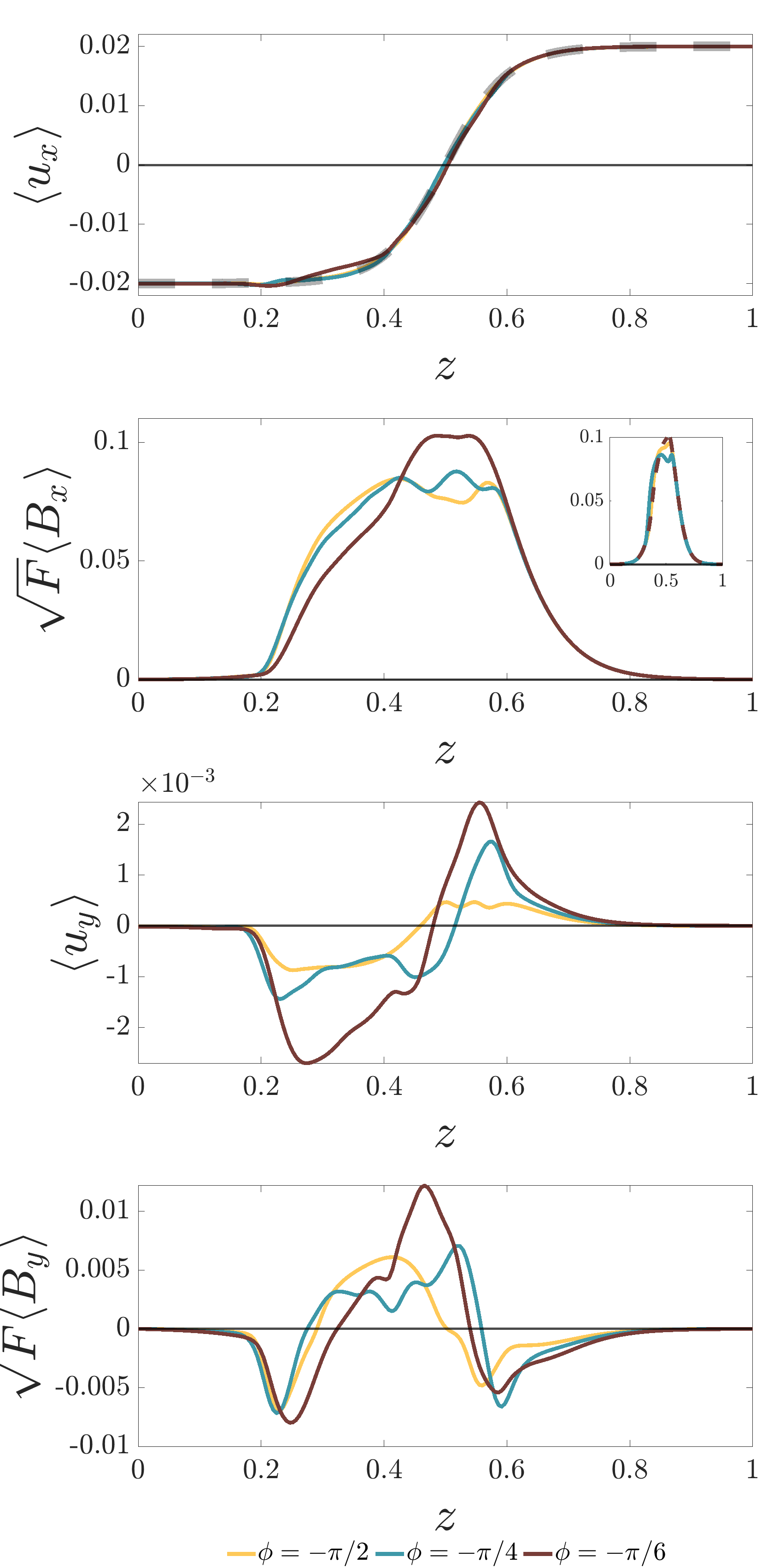}
		\caption{Horizontally averaged vertical profiles for each latitude $\phi\in \{-\pi/2,-\pi/4, -\pi/6\}$ as denoted by the legend, for $\sigma=0.01$. 
        In all cases the profiles are taken at $t\approx 500$ (with additional profiles at $t\approx 440$ for $\sqrt{F}\langle B_x \rangle$ as an insert which further illustrates the evolution of the field).}
		\label{figure_tilted_profiles}
	\end{center}
\end{figure}

To assess the effects of a tilted rotation vector in a more quantitative manner, the mean fields and flows at $t\approx500$ are plotted in Fig.~\ref{figure_tilted_profiles}, which shows snapshots of these quantities at this time.
In all cases, the shear flow profile remains close to the target flow profile, $U_0$, even well into the nonlinear regime.
The secondary $\langle u_y\rangle$ flow remains weak (an order of magnitude smaller than the shear flow in the $x$-direction), but reaches a slightly higher peak value in the low latitude ($\phi=-\pi/6$) case. 
The evolution of $\langle B_x\rangle$ is similar in all cases with only minor quantitative differences being observed (with the most strongly inclined case deviating slightly from the other two). 
As a consequence of the relatively low amplitude of $\langle u_y\rangle$ in all cases, the amplitude of $\langle B_y\rangle$ is generally an order of magnitude smaller than the amplitude of $\langle B_x\rangle$ at a comparable depth. 
Again, the mean field maintains a high degree of alignment with the $x$-axis.

\begin{figure*}
\begin{center}
	\includegraphics[width=\textwidth]{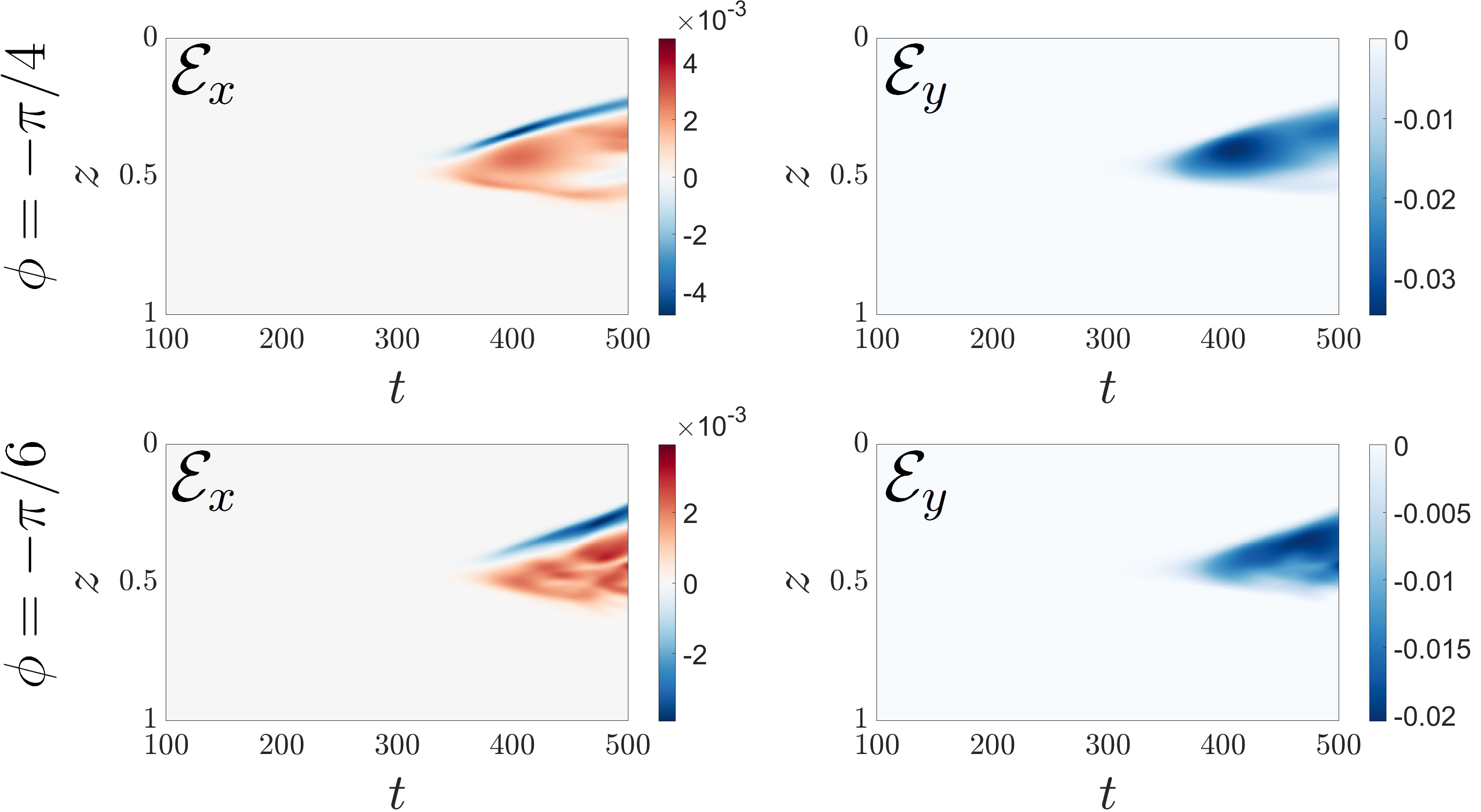}
	\caption{The depth- and time-dependence of $\mathcal{E}_x$ and $\mathcal{E}_y$ for latitudes $\phi \in\{-\pi/4, -\pi/6\}$ in the $\sigma=0.01$ case. 
    These can be compared with the third row of Fig.~\ref{figure_EMF_diff} which shows corresponding quantities for $\phi = -\pi/2$.}
	\label{figure_EMF_tilted}
\end{center}
\end{figure*}

Given that the dynamics seem to be only weakly dependent on the tilt angle, we might expect to see comparable mean EMF profiles. 
The components of the mean EMF for the $\phi=-\pi/4$ and $\phi=-\pi/6$ cases are plotted in Fig.~\ref{figure_EMF_tilted}.
Comparing these profiles with the lower panels of Fig.~\ref{figure_EMF_diff}, we see that these are indeed remarkably similar. 
One small difference is that the maximum amplitude of each component decreases slightly as the latitude is reduced.
Indeed, on symmetry grounds we would expect $\mathcal{E}_x$ to vanish at the equator, i.e.~for $\phi=0$ \citep[e.g.][]{davies_mean_2011}.
Nonetheless, these results demonstrate that a significant EMF can be generated by magnetic buoyancy instability
across a wide range of latitudes.

\section{Discussion and conclusions}

Magnetic buoyancy in the presence of rotation provides a natural alternative mechanism to convection for the ``rise and twist'' effect required for a Parker-like scenario for the solar dynamo  \citep{parker_hydromagnetic_1955,parker_solar_1993}.
The simulations presented in this paper, which built upon those of \citetalias{duguid_shear_driven_2023}, considered the evolution of a shear-generated magnetic layer in a rotating domain, focusing initially upon the simplest case of a vertical rotation vector. 
A shear flow is maintained in the $x$-direction, via the imposition of a body force that balances the viscous and Coriolis terms in the governing equations. 
An initially uniform vertical magnetic field is stretched out in the direction of the flow, producing a magnetic layer that is susceptible to magnetic buoyancy instability.
One of the key limitations of the calculations that were presented in \citetalias{duguid_shear_driven_2023} was that the mean horizontal flow $\langle u_x \rangle$ evolved substantially away from the initial forced shear (due to the dynamical influence of the generated magnetic layer), with significant mean flows also being driven in the $y$-direction.
We were able to resolve this issue by increasing the Prandtl number, which made it more difficult for the Lorentz force to disrupt the initial hydrodynamic force balance, and found only minimal departures from the target flow for $\sigma\in\{0.005,0.01\}$.
This persistent shear flow generated a stronger magnetic layer and hence a more vigorous magnetic buoyancy instability. 
 
We also analysed the resulting mean EMF in the new higher $\sigma$ cases, comparing this analysis with the low $\sigma$ case of \citetalias{duguid_shear_driven_2023}. 
The larger $\sigma$ simulations result in a more complicated time- and depth-dependence of $\mathcal{E}_x$ than that of the low $\sigma$ cases.
In particular, $\mathcal{E}_x$ has dual banded structure with a positive band around the mid-plane that expands in time, above which there is a shallow negative band that moves towards the surface as the simulation progresses.
Crucially, $\mathcal{E}_x$ is of much higher amplitude than in the lower $\sigma$ case that was considered in \citetalias{duguid_shear_driven_2023},
and so is much more likely to play a significant regenerative role in a corresponding dynamo model.
We found that this mean EMF component was largely generated by undular modes of the underlying magnetic buoyancy instability, and this partially accounts for the differences between the lowest Prandtl number case (from \citetalias{duguid_shear_driven_2023}), in which the instability remained largely interchange-like, and the higher Prandtl number calculations that were presented here. 
Undular modes also make the dominant contribution to $\mathcal{E}_y$, which has a simpler structure than $\mathcal{E}_x$ but again reached higher amplitudes in the higher Prandtl number cases. 

Finally, we investigated the effects of varying the inclination of the rotation vector, which corresponds to changing the latitudinal position of the computational domain within the tachocline.
The inclination angle clearly played a dynamical role during the linear development of the buoyancy instability,
but in the nonlinear regime we found only minor quantitative differences between the cases. 
For all values of the inclination angle studied,
we obtained a substantial mean EMF, which is evidence that this regenerative mechanism can operate effectively at the lower latitudes that are of most relevance for the solar dynamo.

Having established computationally accessible conditions under which this system is able to produce a significant mean EMF, whilst also
maintaining a tachocline-like shear flow, the next logical step is to investigate the corresponding dynamo problem (adjusting the initial imposed magnetic field so that there is zero net magnetic flux across the domain).
Based on standard mean-field arguments
\citep{parker_hydromagnetic_1955,moffatt_magnetic_1978,parker_solar_1993,moffatt_self-exciting_2019}
we anticipate that the combination of shear and a significant mean EMF in the direction of the mean field will be conducive to dynamo action.
In fact, these are precisely the ingredients required for a ``dynamo wave'' model,
i.e.~a migratory dynamo similar to that observed in the Sun.
Further work to explore these ideas is underway, and will be presented in a future publication. 

\section*{Acknowledgements}
This work was supported by a Research Project Grant from the Leverhulme Trust (RPG-2020-109). 
CD and TSW were also supported by STCF grants ST/X001083/1 and ST/W001020/1, respectively.
This research made use of the Rocket High Performance Computing service at Newcastle University. 
This work also used the DIRAC Shared Memory Processing system at the University of Cambridge, operated by the COSMOS Project at the Department of Applied Mathematics and Theoretical Physics on behalf of the STFC DiRAC HPC Facility (\href{www.dirac.ac.uk}{www.dirac.ac.uk}). This equipment was funded by BIS National E-infrastructure capital grant ST/J005673/1, STFC capital grant ST/H008586/1, and STFC DiRAC Operations grant ST/K00333X/1. DiRAC is part of the National e-Infrastructure.

\section*{Data availability}
The data underlying this article will be shared on reasonable request to the corresponding author.

\bibliographystyle{mnras}
\bibliography{mybib.bib}



\bsp	
\label{lastpage}
\end{document}